\documentclass[sigconf]{acmart}

\usepackage{algorithm}
\usepackage{algorithmic}
\usepackage{makecell}
\usepackage{subfigure}
\usepackage{multirow}
\usepackage{amsmath}
\usepackage{amsfonts}
\usepackage{adjustbox}

\usepackage{balance}
\usepackage{color}

\AtBeginDocument{%
  \providecommand\BibTeX{{%
    \normalfont B\kern-0.5em{\scshape i\kern-0.25em b}\kern-0.8em\TeX}}}

\copyrightyear{2023}
\acmYear{2023}
\setcopyright{acmlicensed}
\acmConference[SIGIR '23] {Proceedings of the 46th International ACM SIGIR Conference on Research and Development in Information Retrieval}{July 23--27, 2023}{Taipei, Taiwan.}
\acmBooktitle{Proceedings of the 46th International ACM SIGIR Conference on Research and Development in Information Retrieval (SIGIR '23), July 23--27, 2023, Taipei, Taiwan}
\acmPrice{15.00}
\acmISBN{978-1-4503-9408-6/23/07}
\acmDOI{10.1145/3539618.3591845}
\begin{document}

\title{Graph Enhanced BERT for Query Understanding}



\author{Juanhui Li}
\affiliation{%
  \institution{Michigan State University
   \country{USA}}}
\email{lijuanh1@msu.edu}

\author{Wei Zeng}
\affiliation{%
  \institution{Baidu Inc.
   \country{China}}}
\email{zengwei04@baidu.com}

\author{Suqi Cheng}
\affiliation{%
  \institution{Baidu Inc.
   \country{China}}}
\email{chengsuqi@baidu.com}

% \author{Suqi Cheng}
% \affiliation{%
%   \institution{Baidu Inc.
%   \country{China}}}
% \email{chengsuqi@baidu.com}

\author{Yao Ma}
\affiliation{%
  \institution{New Jersey Institute of Technology
   \country{USA}}}
\email{yao.ma@njit.edu}

\author{Jiliang Tang}
\affiliation{%
  \institution{Michigan State University
   \country{USA}}}
\email{tangjili@msu.edu}

\author{Shuaiqiang Wang}
\affiliation{%
  \institution{Baidu Inc.
  \country{China}}}
\email{shqiang.wang@gmail.com}

\author{Dawei Yin}
\affiliation{%
  \institution{Baidu Inc.
  \country{China}}}
\email{yindawei@acm.org}

\renewcommand{\shortauthors}{Juanhui Li et al.}
\begin{abstract}
Query understanding plays a key role in exploring users' search intents. 
% and facilitating users to locate their most desired information. 
However, it is inherently challenging since it needs to capture semantic information from short and ambiguous queries and often requires massive task-specific labeled data. In recent years, pre-trained language models (PLMs) have advanced various natural language processing tasks because they can extract general semantic information from large-scale corpora. 
  However, directly applying them to query understanding is sub-optimal because  existing strategies rarely consider to boost the search performance. 
 On the other hand, search logs contain user clicks between queries and urls that provide rich users' search behavioral information on queries beyond their content.
  Therefore, in this paper, we aim to fill this gap by exploring search logs. 
  In particular, we propose a novel graph-enhanced pre-training framework, GE-BERT, which leverages both query content and the query graph
  % The model is trained on a query graph where nodes are queries and two queries are connected if they lead to clicks on the same urls,
  to capture both semantic information and users' search behavioral information of queries.
  Extensive experiments on offline and online tasks have demonstrated the effectiveness of the proposed framework.
\end{abstract}





\maketitle

\section{Introduction}
\noindent Query understanding~\cite{stoilos2022type} plays a crucial role in information retrieval. It aims to learn the intentions of a search query, and  provides useful information to advance downstream applications such as document retrieval and ranking~\cite{jiang2016learning}. Many types of tasks have been developed to facilitate query understanding such as query classification~\cite{cao2009context,shen2006query}, query matching~\cite{li2014semantic}, and query expansion~\cite{zheng2020bert}. The challenges for these tasks are mainly from two aspects. First, search queries are often short and ambiguous; as a result, it is hard to capture the semantic meanings of queries only relying on their content. 
% Additional information on queries such as domain data (e.g. search log) or external sources (e.g. Knowledge Graph) has been leveraged to address this challenge~\cite{zhao2014tailor}. 
Second, the majority of existing
solutions~\cite{cao2009context,beitzel2005improving} often train models from scratch in a supervised way, which require a large number of task-specific labeled data. However, obtaining such labeled data is labour intensive and usually costs tremendous money. 

Recent years we have witnessed immense efforts in developing pre-trained language models (PLMs)~\cite{shang2019pre, jeong2020context} such as BERT~\cite{DevlinCLT19} and its variants ~\cite{sun2019ernie,yang2019xlnet,liu2019roberta}. These PLMs are trained on large-scale unlabeled corpora where they can extract general semantic information. Thus, they pave a way to help specific NLP tasks even when only a small amount of task-specific labeled data is available. 
% In fact, these PLMs have been proven to significantly improve numerous natural language processing tasks such as text classification~\cite{sun2019fine}, question answering~\cite{yu2020technical}, and single sentence tagging~\cite{DevlinCLT19}. It is evident from recent works that these PLMs can also help tackle aforementioned two challenges in query understanding~\cite{mustar2020using,chen2019bert}. 
However, directly applying these PLMs to query understanding can be sub-optimal. First, the goal of query understanding is to boost the search performance but the current PLMs seldom incorporate this goal into their pre-training strategies. Second, queries are  different from normal natural languages because they are usually short. 

To enhance the PLMs towards query understanding, one natural direction is to design domain-adaptive pre-training strategies with domain data~\cite{zheng2020bert,mustar2020using, chen2019bert,liu2020deep}. The search log is a commonly used domain data for query understanding, which is often denoted as a query-url bipartite click graph~\cite{jiang2016learning}. 
% In this click graph, nodes are sets of queries and urls and an edge is built between a query and a url when there is at least one click between them. 
The graph encodes user search behaviors and provides rich information about queries beyond their content. 
% For example, semantically similar queries are likely to share many clicks in the graph. Therefore, this click graph can greatly help understand users' intents in queries from the users' search behavior perspective. It indeed has been widely utilized to advance many information retrieval tasks~\cite{jiang2016learning, huang2013learning,deng2009entropy}. 
Thus, it is appealing to explore this click graph to advance PLMs. However, BERT and its variants cannot directly process such graph data and dedicated efforts are desired. 

In this paper, we propose a novel graph-based domain adaptive pre-training framework, named Graph Enhanced BERT(GE-BERT), which enjoys both advantages of the PLMs and the domain knowledge, i.e. click graph, for query understanding. 
% In other words, GE-BERT captures not only the semantic information from query content as traditional PLMs, but also the user search behavioral information from the click graph.
To evaluate its effectiveness, we first fine-tune GE-BERT and compare it with representative baselines on two offline tasks, i.e., query classification and query clustering. Then we demonstrate the superiority of the fine-tuned GE-BERT on the online deployment based on the the medical query searching.
\section{The Proposed Framework} 
\label{sec:model}

In this section, we present the details of the proposed framework. Before that, we firstly introduce some key concepts and notations. 
The query graph is built from the query-url bipartite graph~\cite{jiang2016learning}, where two queries are connected if they share at least one url in the bipartite graph.
% Without loss of generality, we only consider the query graph as an unweighted graph in this work. Moreover, it is straightforward to extend the proposed framework by associating weights (e.g., the number of co-clicks by two queries in the bipartite graph) into query connections to represent their strength.
Formally, we denote the query graph as $\mathcal{G}=(\mathcal{V},\mathcal{E})$, where $\mathcal{V} = \{v_1, v_2, ..., v_n\}$ is the set of $|\mathcal{V}| = n$ query nodes, and $\mathcal{E}$ is the set of edges connecting nodes among $\mathcal{V}$. 
Furthermore, the query graph can also be described by an adjacency matrix $\mathbf{A} \in \mathbb{R}^{n \times n}$, where $\mathbf{A}_{ij} = 1$ if the nodes $v_i$ and $v_j$ are connected and $\mathbf{A}_{ij} = 0$ otherwise. Each query node $v_i$ in the graph represents a query $q_i= \{c^i_1,\dots, c^i_t\}$, which consists of a sequence of tokens. Here $c^i_j$ indicates the $j$-th token in $q_i$, and $t$ is the number of tokens.
%in $q_i$. 
% We further summarize the notations in the paper and their descriptions in Table~\ref{table:notation}.

\begin{figure}
  \centering
  \includegraphics[width=0.97\linewidth]{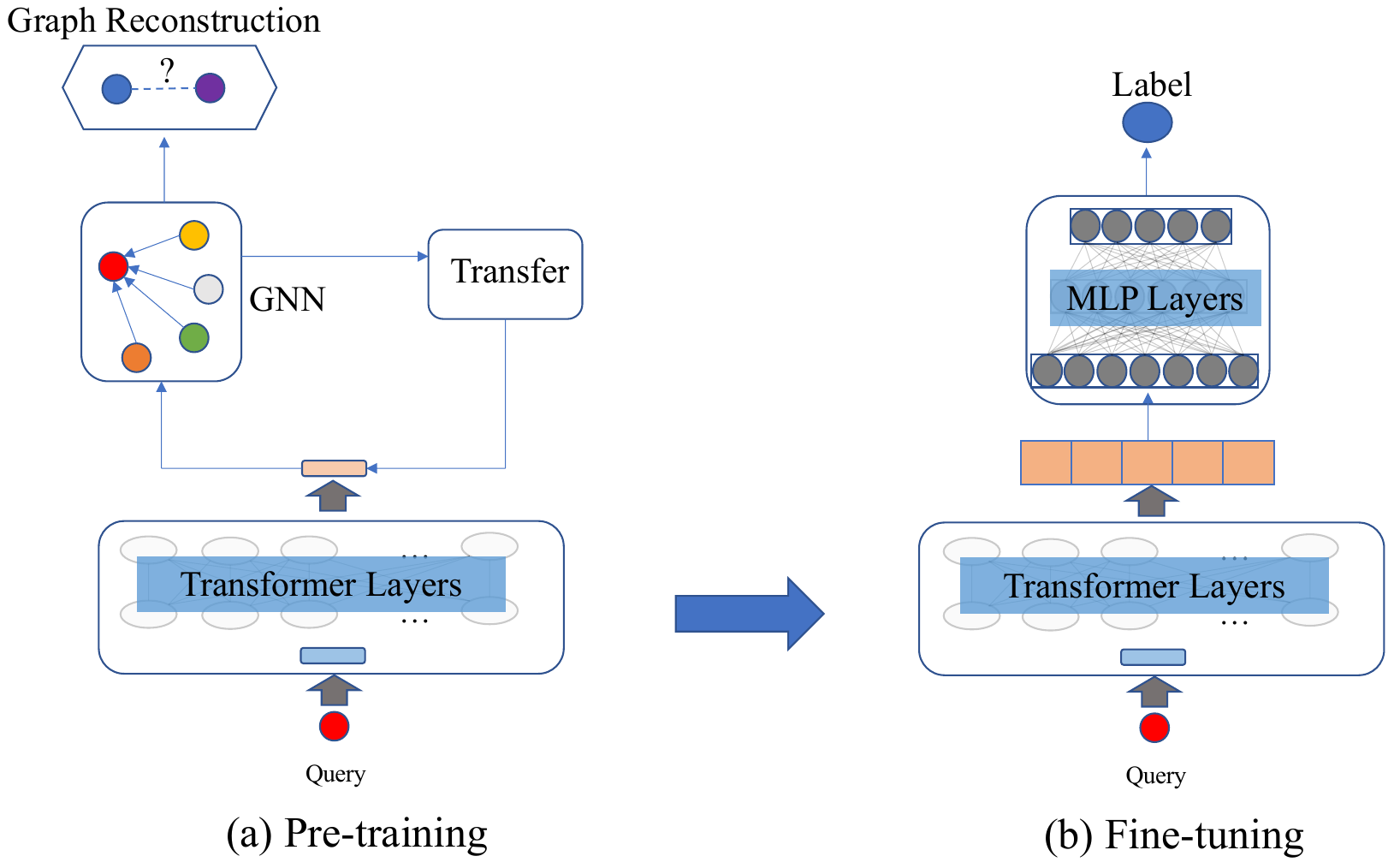}
  \caption{{Illustration of pre-training and fine-tuning process.}}
   %We propose a pre-training framework to train the Transformer module, which then serves as the initialization to build the fine-tuning model.
  \label{fig:framework}
  \vspace{-1.7em}
\end{figure}

Based on the query graph, we propose a pre-training framework GE-BERT to integrate the graph information into BERT, which can be further fine-tuned for downstream query understanding tasks. 
An illustration is shown in \figurename~\ref{fig:framework}.
In particular, the pre-training framework in \figurename~\ref{fig:framework}(a)  consists of three modules -- the Transformer module, the GNN module and the Transfer module. 
The \emph{graph semantics} captured from the GNN module is transferred through the Transfer module to the   Transformer module. An illustration of the downstream task is presented in \figurename~\ref{fig:framework}(b).
% We use various downstream tasks to evaluate the effectiveness of the pre-trained Transformer module in the \textbf{Experiment} section.

% The Transformer module is to extract the \emph{semantic features} from the query content. 
% The GNN module is to yield the \emph{graph semantics} by capturing the query graph information associated with the semantic features. 
% Then the Transfer module transfers the graph semantics to the Transformer module for a better refinement. 
% After pre-training, the Transformer module can capture both the semantic information and the query graph information, and can be fine-tuned for performing downstream tasks. An illustration of the downstream task is presented in \figurename~\ref{fig:framework}(b). We use various downstream tasks to evaluate the effectiveness of the pre-trained Transformer module in the \textbf{Experiment} section. 
Notably, we inject the query graph information into the Transformer module to perform downstream  tasks, instead of using both the Transformer module and the GNN module to build the downstream model
due to two reasons:
% \begin{itemize}
%  \item 
 1) First, the GNN module is only trained on the existing query graph and thus it cannot handle new coming queries without any connection to the existing graph. The Transformer module, on the other hand, has the flexibility to generate embeddings for new queries. 
%   \item 
  2) Second, the complexity of the GNN module for fine-tuning and online deployment is significantly higher than that of the Transformer module, while the Transformer module provides more practical flexibility.
% \end{itemize}
 Next, we introduce details of the three modules in the proposed pre-training framework and discuss pre-training strategies.

\subsection{The Transformer Module}
\label{sec:bert}
In the Transformer module,  we adopt the Transformer encoder of the BERT model to learn representations of queries. 
% Queries are often short, ambiguous, and polysemous, which make it hard to understand queries. We adopt the Transformer structure to learn the representations of queries, and the focused attention mechanism in Transformer could help to disambiguate the polysemous words and homonyms. Thus, to facilitate the training, we utilize the encoder part (which is a Transformer) of the BERT pre-training model.
The encoder is pre-trained on a large scale corpus and thus has strong ability to capture semantic information in natural language.
% which can be adopted to learn good query representations.
The Transformer encoder takes a sequence of tokens as input and outputs a sequence of their corresponding final embeddings, i.e.,
one embedding for each token. 
To feed the queries into the Transformer encoder, we tokenize each query into a sequence of tokens based on the WordPiece method~\cite{wu2016google}. Following the original BERT~\cite{DevlinCLT19}, for each query, we add a special token [CLS] at the beginning of the token sequence; thus the query $q_i$ is denoted as: $q_i= \{[CLS],c^i_1,\dots, c^i_t\}$. 
% Then, for each token in the query, we obtain its input embedding by summing its token embedding and position embedding. Here the token embedding indicates the semantic information of the token while the position embedding captures the position information of the token in the sequence. 
% Both of them are dealt in the same way as in BERT~\cite{DevlinCLT19}. 
We specifically denote the input embedding and output embedding for a token $c^i_j$ in query $q_i$ as ${\bf E}_{c^i_j}$ and ${\bf E}'_{c^i_j}$, respectively. Formally, we summarize the aforementioned process into the following formulation:
\begin{equation}
% \footnotesize
\{\mathbf{E}'_{{CLS}^i}, \mathbf{E}'_{c^i_1}, \dots, {\bf E}'_{c^i_t}\}= \text{Transformer}( \{\mathbf{E}_{{CLS}^i}, \mathbf{E}_{c^i_1}, \dots, {\bf E}_{c^i_t}\}). 
\label{eq:transformer}
\end{equation}
Finally, we use the output embedding of the [CLS] token ${\bf E}'_{{CLS}^i}$ as the final embedding of the entire query.  For convenience, we denote the final embedding output from the Transformer module for query $q_i$ as ${\bf H}_i= {\bf E}'_{{CLS}^i}$.

\subsection{The GNN Module}

To capture the graph information in the query graph, we propose the GNN module which takes the query graph as input. 
The node's initial embedding is represented by the output embedding from the Transformer module for the specific query. 
The GNN module consists of several GNN layers, where each layer updates the node embedding by aggregating information from its neighborhood.
%defined according to the graph structure. 
Specifically, the GNN model is able to capture the $K$-hop neighborhood information after $K$ layers. Formally, the $k$-th GNN layer can be generally defined as follows:
\begin{equation}
    \tilde{\bf{H}}^{k} = \text{GNN}({\bf A}, \tilde{\bf H}^{k-1})
\end{equation}
%\jt{please define $\tilde{\bf H}^{k-1}$ in the above formulation}
where $\text{GNN}(\cdot)$ denotes a general GNN layer, ${\bf A}$ is the adjacency matrix and $\tilde{\mathbf{H}}^{k} \in  \mathbb{R}^{n\times d_{k}}$ is the output embedding of all nodes in the $k$-th layer. 
We use $\tilde{\mathbf{H}}_{i} \in  \mathbb{R}^{d_{K}}$ to denote the final output  embedding of the node $v_i$ after a total of $K$ iterations, i.e., $\tilde{\bf H}_i = \tilde{\bf H}_i^K$.
Specifically,  the node embedding is initialized with the output from the Transformer module, i.e., $\tilde{\mathbf{H}}^{0}_{i}=\mathbf{H}_{i}$.

In this paper, we adopt two representative GNN models, i.e., GCN~\cite{KipfW17} and GAT~\cite{VelickovicCCRLB18}, to capture the query graph information.
To preserve the graph information and train the GNN parameters, we use the output node embeddings from the GNN module to reconstruct the query graph. In particular, given two nodes ${v}_i$ and ${v}_j$,
we define a probability score $ s_{ij}=\text{sigmoid} (\tilde{\mathbf{H}}_{i} \cdot \tilde{\mathbf{H}}_{j})$ to predict if they have a link in the query graph
% we utilize their embeddings $\tilde{\bf H}_i$ and $\tilde{\bf H}_j$ to predict if they have a link in the query graph.
% The probability score $s_{ij}$ of nodes $v_i$ and $v_j$ with a link is modeled as $ s_{ij}=\text{sigmoid} (\tilde{\mathbf{H}}_{i} \cdot \tilde{\mathbf{H}}_{j})$
% \begin{align}
%     s_{ij}=\text{sigmoid} (\tilde{\mathbf{H}}_{i} \cdot \tilde{\mathbf{H}}_{j})
% \end{align}
\noindent where $\text{sigmoid}(\cdot)$ is the sigmoid function. 
The loss  of the graph reconstruction is defined as follows:
\begin{equation}
\label{eq:GNN_lp}
\mathcal{L}_{GNN} = -\sum_i \sum_{
% (i, j) \in \mathcal{V}
j \in( \mathcal{N}_i\cup  \mathcal{N}'_i)
} \mathbf{A}_{ij}\log s_{ij} +(1-\mathbf{A}_{ij}) \log (1 - s_{ij})
\end{equation}
% where $y_{ij} = 1$ indicates that nodes ${v}_i$ and ${v}_j$ are connected in the query graph, and $y_{ij} = 0$ otherwise. 
where $\mathcal{N}_i$ is the set of neighbors of $v_i$, 
$  \mathcal{N}'_i$ is the set of negative samples that are not connected with node $v_i$, and  $| \mathcal{N}'_i| = | \mathcal{N}_i| $.

\subsection{The Transfer Module}

The Transfer module is proposed to fuse the graph information captured by the GNN module to the Transformer module. In the literature, minimizing the KL divergence between prior and posterior distributions is demonstrated to be an effective way for knowledge transfer~\cite{zhan2020user,lian2019learning}. 
% In this module, we would like to minimize the gap between the GNN module and the Transformer module.
% To measure this gap, we use KL-divergence which constructs the distribution divergence between the prediction results of these two modules.
% Next, we give more details of this strategy.
Inspired by this, we define the prior distribution based on the  Transformer module since it only utilizes query content: $p(\mathbf{A}_{ij}|\mathbf{H}_{i}, \mathbf{H}_{j}) = {sigmoid(\mathbf{H}_{i} \cdot \mathbf{H}_{j})}$.
% we treat the conditional probability distribution over the query graph edges derived from the Transformer module as a prior distribution, since it only utilizes query content, and treat the one obtained from the GNN module as a posterior distribution, for it involves additional information from user's search behaviors. Then we minimize the KL divergence between the two distributions. By approximating the prior distribution to the posterior distribution, the graph information captured by the GNN module can be transferred into the Transformer module.
% In the prior distribution, the conditional probability distribution over a query graph edge is defined as follows:
% \begin{equation}
% p(y_{ij}|\mathbf{H}_{i}, \mathbf{H}_{j}) = {sigmoid(\mathbf{H}_{i} \cdot \mathbf{H}_{j})}
% \end{equation}
% In the posterior distribution, the conditional probability distribution over a query graph edge is defined as follows
The posterior distribution is obtained from the GNN module as it involves additional information from users' search behaviors: $\tilde{p}(\mathbf{A}_{ij}|\tilde{\mathbf{H}}_{i}, \tilde{\mathbf{H}}_{j}) = sigmoid(\tilde{\mathbf{H}}_{i} \cdot \tilde{\mathbf{H}}_{j})$.
% \begin{equation}
% \tilde{p}(y_{ij}|\tilde{\mathbf{H}}_{i}, \tilde{\mathbf{H}}_{j}) = sigmoid(\tilde{\mathbf{H}}_{i} \cdot \tilde{\mathbf{H}}_{j})
% \end{equation}
The loss to minimize the distance between these two distributions is defined as follows:

{\small
\begin{align}
\nonumber \mathcal{L}_{KL} = \sum_i \sum_{
% (i, j) \in \mathcal{V}
j \in( \mathcal{N}_i\cup  \mathcal{N}'_i)
} 
& \bigg(\tilde{p}(\mathbf{A}_{ij}|\tilde{\mathbf{H}}_{i}, \tilde{\mathbf{H}}_{j}) \log\frac{\tilde{p}(\mathbf{A}_{ij}|\tilde{\mathbf{H}}_{i}, \tilde{\mathbf{H}}_{j})}{p(\mathbf{A}_{ij}|\mathbf{H}_{i}, \mathbf{H}_{j}) }\\
\nonumber
&+ (1-\tilde{p}(\mathbf{A}_{ij}|\tilde{\mathbf{H}}_{i}, \tilde{\mathbf{H}}_{j})) \log \frac{1-\tilde{p}(\mathbf{A}_{ij}|\tilde{\mathbf{H}}_{i}, \tilde{\mathbf{H}}_{j})}{1-p(\mathbf{A}_{ij}|\mathbf{H}_{i}, \mathbf{H}_{j})}\bigg)
\label{eq:KL}
\end{align}
}
The prior and posterior distributions essentially predict if two nodes are connected.
% based on embeddings from the Transformer module and the GNN module, respectively. 
Optimizing this loss potentially drives the Transformer module to achieve similar graph reconstruction result with the GNN module, i.e., the Transformer module could naturally capture the graph information guided by the GNN module.

\subsection{Pre-training Strategies}
\label{sec: strategies}

We propose two strategies for the pre-training process, i.e., stage-by-stage and joint training.
1) In the \textbf{stage-by-stage strategy}, we separate the training processes of the Transformer module and the GNN module. In the stage of training the GNN module, we fix the parameters of the Transformer module and update the GNN module by the loss  $\mathcal{L}_{GNN}$. In the stage of training the Transformer Module,  we fix the GNN module and transfer the information captured by the GNN module into the Transformer module by the loss  $\mathcal{L}_{KL}$.  
2) In the \textbf{joint training strategy}, the proposed modules are trained simultaneously. The loss function is defined by combining two losses: $\mathcal{L}_{J} = \mathcal{L}_{GNN} + \lambda \mathcal{L}_{KL}$, where $\lambda$ is a pre-defined parameter to balance the two losses. For both strategies, the trained Transformer module can be used to build models for downstream tasks.

\section{Experiment}
\label{sec:exp}

In this section, we aim to validate the effectiveness of GE-BERT. We first fine-tune it on two {\bf offline} query understanding tasks, i.e., query classification and query matching tasks. 
Then, we further demonstrate the effectiveness of the fine-tuned model in the {\bf online} search system based on the medical query searching.
% and online experiments on the query understanding tasks. We first show that our method can boost the performance on two offline tasks, i.e., query classification and query matching tasks, then we further show that our method can advance the online search performance by taking the medical query classification as an example.

\subsection{Datasets}

% Statistics of datasets in the pre-training, query classification and query matching are shown in Table~\ref{table:data}. 
We use different datasets according to the tasks.  1) For the \textbf{pre-training},  the model is trained on the query graph which is built on a large query log data generated by the Baidu search engine\footnote{https://www.baidu.com/} with millions of users. We collected the initial search log data within a week to construct 
the query graph which consists of $92,413,932$ query nodes and $957,210,956$ edges. 
% The average degree of nodes is $21$. 
% For the two offline tasks, we sample datasets from the search log of the search engine, and manually label the samples. 
2) In the {\bf offline query classification} task, we predict the label for the input query.
Each query is associated with a label such as music and medical care. 
There are $100,887$ queries with $30$ classes in total. 
3) In the {\bf offline  query matching} task, we predict if the input query pairs have similar semantic meaning or not. Each query pair is associated with a label which is $1$ when the two queries have similar semantic meanings, and $0$ otherwise.
% An input query pair has the label $1$ when the two queries have similar semantic meanings, otherwise, the label is $0$.
There are $108,615$ query pairs with $2$ classes in total. 
%  The statistics of the datasets is shown in Table \ref{table:data}.
 We adopt $60\%, 20\%, 20\%$  for the training/validation/test split for offline  tasks.

\subsection{Offline tasks}

\subsubsection{Baselines and Model Variants}
We consider two baselines based on the BERT model~\cite{DevlinCLT19}.
The first one directly uses BERT model in the downstream task, denoted as \textbf{BERT}. The second one further trains on the query graph under the graph reconstruction task, and we denote it as \textbf{BERT+Q}. Namely, different from the \textbf{BERT}, \textbf{BERT+Q} further trains the BERT parameters on the query graph by the loss function which has the same form as Eq.(\ref{eq:GNN_lp}).
\textbf{BERT+Q} is a baseline to evaluate the effectiveness of the GNN module to capture query graph information.  Two variants are
\textbf{GE-BERT-J}  which jointly trains all proposed modules and \textbf{GE-BERT-S} which uses the stage-by-stage pre-training strategy. 

% We give more details of these two representative baselines together with model variants as follows:
% \begin{itemize}
%     \item  \textbf{BERT}: 
   
%     It's a baseline where we directly adopt the traditional bert-base-chinese\footnote{https://github.com/google-research/bert}
%     model to do the fine-tuning without pre-training on the query graph.
%     \item \textbf{BERT+Q}:  It's a baseline directly built upon \textbf{BERT}, which is further pre-trained on the query graph under the graph reconstruction task: the loss function has the same form as Eq.(\ref{eq:GNN_lp}) and the probability score $s_{ij}$ is generated based on the embeddings from the BERT rather than the GNN module.
   
%     \item  \textbf{GE-BERT-J}: A variant of the proposed model which jointly trains the all proposed modules.
%      \item  \textbf{GE-BERT-S}: A variant of the proposed model which uses the stage-by-stage pre-training strategy. 

% \end{itemize}

% Note that there are two  GNN models to extract the query graph information, i.e., GCN and GAT.
% We will report the performance of the variants of the proposed framework under both models. 

\subsubsection{Experimental Settings}
For pre-training tasks, the number of stages in GE-BERT-S is set to be 4, and $\lambda$ in GE-BERT-J is set to be 1. 
For the two downstream tasks, we utilize the Transformer module to build models. 
% More specifically, we include an additional Multi-Layer Perceptron (MLP) model to take the query embeddings generated from the Transformer as input and produce the corresponding output for the downstream tasks. 
Specifically, for the query classification task, given a query, we feed the query embedding  from the Transformer Module into a Multi-Layer Perceptron (MLP) for classification. For the query matching task, given a query pair, we feed their embeddings into a MLP and then
calculate the inner product followed by the sigmoid function to generate the score for final prediction.

\subsubsection{Performance }
% We adopt two different training/validation/test splits, 
% i.e., $80\%, 10\%, 10\%$
The experiment results are shown in Table~\ref{table:query_cls_311} and Table~\ref{table:query_matching_311} respectively.
% Note that the  performance is assessed by accuracy (or ACC), Precision, Recall and F1. 
% We present the results with split of  $80\%, 10\%, 10\%$ in  Table~\ref{table:query_cls_811} and results with split of $60\%, 20\%, 20\%$ in  Table~\ref{table:query_cls_311}.
% Note that the classification performance is assessed by accuracy (or ACC), Precision, Recall and F1. From Table~\ref{table:query_cls_811} and
From these tables,  we can make the following observations: 1) BERT+Q outperforms BERT consistently which demonstrates that incorporating query graph information can enhance Transformer to perform better. 2) In general, the proposed GE-BERT variants achieve better performance than BERT+Q, which indicates that the GNN module can capture the query graph information more effectively than pre-training with simple graph reconstruction task, which facilitates the performance. It also suggests  that KL-divergence can indeed facilitate the process of transfer the information from the GNN module to the Transformer module. 

\begin{table}[]
\caption{Performance of the offline query classification task. 
% The split of the dataset is $60\%, 20\%, 20\%$ for training,validation and test, respectively.
}
\label{table:query_cls_311}
\centering
\small
\begin{adjustbox}{width =0.45 \textwidth}
\begin{tabular}{cc|ccccc}
 \toprule
\multicolumn{2}{c|}{\multirow{1}{*}{Methods}} & \multirow{1}{*}{ACC}& \multirow{1}{*}{Precision} & \multirow{1}{*}{Recall} & \multirow{1}{*}{F1}\\ 
% &&&&&&&
%\cr
\midrule
\multicolumn{2}{c|}{BERT}& 0.6728 & 0.5941& 0.5604 & 0.5722 \\
\multicolumn{2}{c|}{BERT+Q}& 0.6862&0.6162& 0.5708 & 0.5863 \\ 
\midrule

\multicolumn{1}{c}{\multirow{2}{*}{{GAT-based}}}&{{GE-BERT-J}}&0.6897 & 0.6215& 0.5702 & 0.5874 \\

% \multicolumn{1}{c|}{}&\multicolumn{1}{c|}{}&cos&  0.6710& 0.6085& 0.5465 &0.5660  \\
% \cline{2-7}
\multicolumn{1}{c}{}&{GE-BERT-S}& {0.7033} & \textbf{0.6308}& 0.5835  & 0.5997 \\
% \multicolumn{1}{c|}{}&\multicolumn{1}{c|}{}& cos& \bf{0.7062} & 0.6360& \bf{0.6099} & \bf{0.6162}\\
\midrule
\multicolumn{1}{c}{\multirow{2}{*}{{GCN-based}}}&{{GE-BERT-J}}&0.6945&0.6292 &0.5758 &0.5923  \\

% \multicolumn{1}{c|}{}&\multicolumn{1}{c|}{}&cos& 0.6750 &0.6182 &0.5548  & 0.5765\\
% \cline{2-7}
\multicolumn{1}{c}{}&{{GE-BERT-S}}&\textbf{0.7034}  & 0.6227& \textbf{0.5978} & \textbf{0.6034} \\
% \multicolumn{1}{c|}{}&\multicolumn{1}{c|}{}& cos& 0.7028 & \textbf{0.6375}&0.5874 &0.6029 \\

\bottomrule
\end{tabular}
\end{adjustbox}
\vspace*{-6mm}
\end{table}

\begin{table}[]
\caption{Performance of the offline query matching task. 
% The split of the dataset is $60\%, 20\%, 20\%$ for training,validation and test, respectively.
}
\label{table:query_matching_311}
\centering
\small
\begin{adjustbox}{width =0.45 \textwidth}
\begin{tabular}{cc|ccccc}
 \toprule
\multicolumn{2}{c|}{\multirow{1}{*}{Methods}} & \multirow{1}{*}{ACC}& \multirow{1}{*}{Precision} & \multirow{1}{*}{Recall} & \multirow{1}{*}{F1}\\ 
% &&&&&&&
\midrule
\multicolumn{2}{c|}{BERT}& 0.6232 &0.7825 & 0.6282 &0.5661  \\
\multicolumn{2}{c|}{BERT+Q}& 0.6254& 0.7842& 0.6303 & 0.5691  \\ 
\midrule

\multicolumn{1}{c}{\multirow{2}{*}{{GAT-based}}}&{{GE-BERT-J}}& 0.6388 &0.7887 & 0.6436 &0.5891  \\

% \multicolumn{1}{c|}{}&\multicolumn{1}{c|}{}&cos&  0.6391&0.7888 &0.6438  &0.5894  \\
% \cline{2-7}
\multicolumn{1}{c}{}&{{GE-BERT-S}}& 0.6370 & 0.7881& 0.6418  &0.5865  \\
% \multicolumn{1}{c|}{}&\multicolumn{1}{c|}{}& cos& 0.6212 &0.7829 & 0.6262 &0.5629 \\

\midrule
\multicolumn{1}{c}{\multirow{2}{*}{{GCN-based}}}&{{GE-BERT-J}}&\textbf{0.6425} & \textbf{0.7899}& \textbf{0.6472} & \textbf{0.5944} \\

% \multicolumn{1}{c|}{}&\multicolumn{1}{c|}{}&cos& 0.6350 & 0.7874& 0.6397 & 0.5834 \\
% \cline{2-7}
\multicolumn{1}{c}{}&{{GE-BERT-S}}&0.6312  & 0.7861& 0.6317  &0.5779  \\
% \multicolumn{1}{c|}{}&\multicolumn{1}{c|}{}& cos& 0.6202 & 0.7825&  0.6252&0.5612 \\

\bottomrule
\end{tabular}
\end{adjustbox}
\vspace*{-3mm}
\end{table}

\subsection{Online task}

% In this section, we conduct an online A/B testing to validate the effectiveness of our models in Baidu search engine. Specifically, the A/B testing is based on the final ranked results for medical query searching, which is an important search scenario and has specific search strategies depended on medical query classification in the Baidu search engine. This means, whether a query is judged to be a medical query leads to different ranked results in the search engine.

In this subsection, we conduct an online A/B testing to validate the effectiveness of our models in the Baidu search engine. It is based on the final ranked results for medical query searching, which is an important search scenario in the Baidu search engine. A crucial step for the medical query searching is to identify medical queries first. To achieve this goal, we treat it as a classification task and directly utilize the {\bf fine-tuned} model (including the MLP layer) from the offline query classification task to perform the classification. In particular, given an input query, the fine-tuned model is able to tell whether it is a medical query or not. We have designed many strategies for medical queries to facilitate their ranking performance. Thus, 
different fine-tuned  models might have different classification performance, which leads to different final ranked results on the search engine.  
We evaluate the model performance by comparing their ranked results. Due to the high cost of deploying multiple models, we evaluate the fine-tuned GE-BERT-S with GCN  ({\bf GE-BERT-S+F})  and make a comparison with the fine-tuned BERT+Q model ({\bf BERT+Q+F}) 

\subsubsection{Experimental Settings and Result}

% we predict if the query is categorized as medical care or not. The query has label $1$ when it's related to medical care, otherwise, it has label $0$. There are  50,000 queries which are randomly sampled from the search engine. }

%\jh{
% To investigate the effectiveness of GE-BERT on the real-world online environment, we deploy the model on the search engine to perform the medical query classification.  Due to the high cost of deploying multiple models, we evaluate  the  GE-BERT-S with  GCN which consistently achieves good performance on the offline tasks, and make a comparison with the BERT+Q model. 
%We randomly sample $50,000$ queries from the search log and we apply GE-BERT-S+F and BERT+Q+F to make a prediction. Then we obtain $6268$ queries which are predicted as medical queries by as least one model from GE-BERT-S+F  and BERT+Q+F. Among them, there are $5304$ queries that they make the same prediction. To compare the performance of GE-BERT-S+F  and BERT+Q+F, we compare the ranked results on the $6268-5304 = 964$ queries that they make different predictions, i.e., the difference ratio is $964/6268=15.38\%$. Further we use the search engine to obtain the ranked results for these $964$ medical queries. }

We randomly sample a certain number of queries from the search log, and we apply GE-BERT-S+F and BERT+Q+F to identify the medical queries. Then we obtain $6,268$ queries which are predicted as medical queries by as least one of the two models. Among these queries, they make different predictions on $964$ queries, i.e., the difference ratio is $964/6268=15.38\%$. 

To compare the performance of GE-BERT-S+F and BERT+Q+F online, we compare their ranked results  obtained from the search engine for these $964$ queries, and use the (Good vs. Same vs. Bad) GSB~\cite{zou2021pre} as the metric. More specifically, given a query, the annotators are provided with a pair $(result_1, result_2)$ where $result_1$ is the ranked result from  GE-BERT-S+F and $result_2$ is the ranked result  from BERT+Q+F. The result consists of multiple URLs. The annotators who  don't know which model the result is generated are asked to rate the result independently: Good ($result_1$ is better), Bad ($result_2$ is better), and Same
(they are equally good or bad) by considering the relevance between the ranked result and the given query. To quantify the
human evaluation, the $\Delta$GSB is defined by  combining these three indicators: 
\begin{align}
    \Delta GSB = \frac{\# \text{Good} - \# \text{Bad} }{\# \text{Good} + \# \text{Same} + \# \text{Bad}}
\end{align}
The statistic for computing the $ \Delta GSB$ is shown in Table~\ref{table:online}. We can observe that  GE-BERT-S+F brings in substantial improvement and the advantage of GSB is $3.20\%$, which shows the superiority of our model to advance the online medical query searching.

% \jh{**how to compute the GSB because I don't know the \#same here?**}

% \jh{**Not so sure if saying "deploy on the search engine" is right**}

% \jh{**I'm confused about the cases that wei  said we can use to show our good performance. Although we have some cases showing our model have the right prediction,  seems it's hard to explain why our model is right by combing the graph, and why the baseline is wrong. Maybe we can discuss later.**}

\begin{table}[]
\caption{Results of the online A/B testing. 
% The split of the dataset is $60\%, 20\%, 20\%$ for training,validation and test, respectively.
}
\label{table:online}
\centering
\small
\begin{tabular}{ccccc}
 \toprule
% \multicolumn{2}{c|}{\multirow{1}{*}{Methods}} & \multirow{1}{*}{ACC}& \multirow{1}{*}{Precision} & \multirow{1}{*}{Recall} & \multirow{1}{*}{F1}\\ 
 diff\_ratio &\# Good & \# Same & \# Bad & $\Delta GSB$ \\
% &&&&&&&
\midrule
15.38\% & 57& 881 &26 &3.20\% \\

\bottomrule
\end{tabular}
\vspace*{-2mm}
\end{table}
\vspace*{-1mm}
\begin{table}[]
\caption{Case study of  query classification for the online task. 
% The split of the dataset is $60\%, 20\%, 20\%$ for training,validation and test, respectively.
}
\label{table:case_online}
\centering
\small
\begin{adjustbox}{width =0.45 \textwidth}
\begin{tabular}{cccc}
 \toprule
% \multicolumn{2}{c|}{\multirow{1}{*}{Methods}} & \multirow{1}{*}{ACC}& \multirow{1}{*}{Precision} & \multirow{1}{*}{Recall} & \multirow{1}{*}{F1}\\ 
Query& Label & GE-BERT-S+F & BERT+Q+F \\
% &&&&&&&
\midrule
        What's the weight of Nini & people& people & medical care \\
\midrule
% \makecell[l]{How to save children that\\ are addicted to video games} &games & games  &medical care\\
\makecell[l]{Dead Cells the fisherman} &games & games  &medical care\\
\midrule
\makecell[l]{Which eyedrops is good for \\long-term use for students} & medical care & medical care  &  product\\ 
\midrule
\makecell[l]{What are the pros and cons \\ of drinking soy milk} & medical care  & medical care &  parenting\\ 
\bottomrule
\end{tabular}
\end{adjustbox}
\vspace*{-5mm}
\end{table}

\subsubsection{Case Study}
In this section, we further analyze the fine-tuned model  by practical cases in query classification.
They are presented in Table~\ref{table:case_online} where the second column is the true label of the query, the third and the forth column are predicted labels from GE-BERT-S+F  and  BERT+Q+F respectively. 
We have the following observations: 
1) For the first two queries, BERT+Q+F wrongly predicts them as medical queries. 
The potential reason might come from two misleading words \emph{weight} and \emph{Cells} because they are related to medical care.
However, the first query is more about the word \emph{Nini} (a famous actress in China) and \emph{Dead Cells} in the second query is a video game.
2) For the third and the forth query, BERT+Q+F wrongly predicts them as other categories instead of the medical query.  
However, the word \emph{eyedrops} in the third query is not simply a product but more related to the medical care. In the forth query, the word \emph{milk} might be misleading so BERT+Q+F predicts it as  \emph{parenting}. In fact, the user cares more about the function of the soy milk so it's a medical query. 
Generally, these queries are difficult ones containing the misleading words. GE-BERT-S+F is able to make the right predictions. The potential reason is that the GNN utilizes the information of neighboring queries which can help to distinguish the misleading words.

\section{Conclusion}
\label{sec:conclusion}
In this paper, we introduce GE-BERT, a novel pre-training framework preserving both the semantic information and query graph information, which is further fine-tuned in the query understanding tasks. Specifically, we devise three modules: a Transformer module for extracting the semantic information of query content, a GNN module for capturing the query graph information, and a Transfer module to transfer the graph information from the GNN module to the Transformer module. To train the GE-BERT, two  strategies are proposed: stage-by-stage strategy by separating the training of the Transformer module and the GNN module, and joint training by training all modules simultaneously. Comprehensive experiments in offline and online tasks demonstrate the superiority of GE-BERT. 
% compared with the state-of-the-art methods. 

\section{Acknowledgements}
This research is supported by the National Science Foundation (NSF) under grant numbers CNS1815636, IIS1845081, IIS1928278, IIS1955285, IIS2212032, IIS2212144, IOS2107215, IOS2035472, and IIS2153326, the Army Research Office (ARO) under grant number W911NF-21-1-0198, the Home Depot, Cisco Systems Inc, Amazon Faculty Award, Johnson\&Johnson and SNAP.

\section{Presenter and Company Biography}

\textbf{Presenter}: Juanhui Li is a PhD student in computer science at Michigan State University. She's mainly interested in the graph mining including graph neural networks, link prediction and knowledge graph.

% \section{Company Biography}
 \noindent \textbf{Company}: Baidu Inc. is one of the largest AI and internet companies in the world which is founded as a search engine platform in 2000. Baidu provides full AI stack services including infrastructure consisting of AI chips, deep learning framework and open AI platform to facilitate a wide range of applications and services.
\bibliographystyle{ACM-Reference-Format}
\balance
\bibliography{sample-base}

\end{document}